\documentclass{article}

\usepackage{arxiv}
\usepackage{xspace}
\usepackage{adjustbox}
\usepackage{amsmath,amssymb,amsfonts}
\usepackage{algorithmic}
\usepackage{graphicx}
\usepackage{textcomp}
\usepackage[table,xcdraw]{xcolor}
\usepackage{booktabs}
\usepackage{listings}
\usepackage{multirow}
\usepackage{enumitem}
 \usepackage{url}

\makeatletter
\let\old@lstKV@SwitchCases\lstKV@SwitchCases
\def\lstKV@SwitchCases#1#2#3{}
\makeatother
\usepackage{lstlinebgrd}
\makeatletter
\let\lstKV@SwitchCases\old@lstKV@SwitchCases

\lst@Key{numbers}{none}{%
    \def\lst@PlaceNumber{\lst@linebgrd}%
    \lstKV@SwitchCases{#1}%
    {none:\\%
     left:\def\lst@PlaceNumber{\llap{\normalfont
                \lst@numberstyle{\thelstnumber}\kern\lst@numbersep}\lst@linebgrd}\\%
     right:\def\lst@PlaceNumber{\rlap{\normalfont
                \kern\linewidth \kern\lst@numbersep
                \lst@numberstyle{\thelstnumber}}\lst@linebgrd}%
    }{\PackageError{Listings}{Numbers #1 unknown}\@ehc}}
\makeatother

\title{ENCORE: \underline{En}semble Learning using \underline{Co}nvolution Neural Machine Translation for Automatic Program \underline{Re}pair}

\author{
  Thibaud Lutellier \\
  University of Waterloo\\
  Canada\\
  \texttt{tlutelli@uwaterloo.ca} \\
   \And
 Lawrence Pang \\
  University of Waterloo\\
  Canada\\
  \texttt{lypang@edu.uwaterloo.ca} \\
   \And
 Viet Hung Pham \\
  University of Waterloo\\
  Canada\\
  \texttt{hvpham@uwaterloo.ca} \\
   \And
 Moshi Wei \\
  University of Waterloo\\
  Canada\\
  \texttt{m44wei@uwaterloo.ca} \\
   \And
 Lin Tan \\
  Purdue University\\
  United States\\
  \texttt{lintan@purdue.edu} \\
}

\begin{document}
\maketitle

\newcommand{\bugsearchspacedefectstotal}{395\xspace}
\newcommand{\bugsearchspacequixbugtotal}{39\xspace}
\newcommand{\bugsearchspacegenprogtotal}{69\xspace}
\newcommand{\bugsearchspacedefects}{261\xspace}
\newcommand{\bugsearchspacequixbug}{32\xspace}
\newcommand{\bugsearchspacegenprog}{21\xspace}
\newcommand{\bugsearchspacedefectspercent}{66\% (77\% with oov)\xspace}
\newcommand{\bugsearchspacequixbugpercent}{82\% (100\% with oov)\xspace }
\newcommand{\bugsearchspacegenprogpercent}{30\% (77\% with oov: solution: add the target program in the embedding would help)\xspace}

\newcommand{\djfix}{28\xspace}
\newcommand{\djfirst}{20\xspace}

\newcommand{\quixfix}{14\xspace}
\newcommand{\quixfirst}{13\xspace}
\newcommand{\pythonfix}{17\xspace}
\newcommand{\cppfix}{5\xspace}
\newcommand{\jsfix}{3\xspace}

\newcommand{\alljava}{42\xspace}
\newcommand{\uniquejava}{16\xspace}
\newcommand{\allfirstjava}{33\xspace}
\newcommand{\uniquedj}{6\xspace}
\newcommand{\uniquefirstdj}{4\xspace}

\newcommand{\alllanguage}{67\xspace}

\newcommand{\trainingdata}{1,159,502\xspace}
\newcommand{\trainingdatacpp}{3,599,472\xspace}
\newcommand{\trainingdatapython}{711,091\xspace}
\newcommand{\trainingdatajavascript}{831,711\xspace}

\newcommand{\tokenjava}{24.3\xspace}
\newcommand{\tokencpp}{17.0\xspace}
\newcommand{\tokenpython}{21.2\xspace}
\newcommand{\tokenjavascript}{17.8\xspace}

\newcommand{\numGitHubProj}{1,000\xspace}
\newcommand{\gv}{G\&V\xspace}
\newcommand{\tool}{ENCORE\xspace}

\newcommand{\javEmbFile}{492,000\xspace}
\newcommand{\javVocSize}{44,109\xspace}

\begin{abstract}
Automated generate-and-validate (\gv) program repair techniques  
typically rely on hard-coded rules, 
only fix bugs following specific patterns, and are hard to adapt to different programming
languages. 

We propose \tool, a new \gv technique, which uses ensemble learning on 
convolutional neural machine translation (NMT)
models to automatically fix bugs in multiple programming languages.
We take advantage of the randomness in hyper-parameter tuning to 
build multiple models that fix different bugs and combine them using
ensemble learning. This new convolutional NMT approach 
outperforms the standard long short-term memory (LSTM) approach used in previous work,
as it better captures both local and long-distance connections between
tokens.

Our evaluation on two popular benchmarks, Defects4J and Quix\-Bugs, shows that \tool{} fixed \alljava bugs, including \uniquejava
that have not been fixed
by existing techniques.
In addition, \tool is the first \gv repair technique to be
applied to four popular programming languages (Java, C++, Python, and JavaScript), 
fixing a total of \alllanguage bugs across five benchmarks.

\end{abstract}

\section{Introduction}

To improve software reliability and increase engineering productivity, 
researchers have developed many approaches to fix software bugs automatically. 
One of the main approaches for automatic program repair 
is the G\&V 
method~\cite{le2012genprog, long2015staged, durieux2016dynamoth, yang2017better, xuan2017nopol, saha2017elixir}. 
First, candidate patches are generated using a set of transformations or 
mutations  (e.g., deleting a line and adding a clause). 
Second, these candidates are ranked and validated by compiling and running 
a given test suite. The G\&V tool returns the highest ranked fix that compiles 
and passes fault-revealing test cases in the test suite.

While G\&V techniques successfully fixed bugs in different data\-sets, a recent study~\cite{long2016analysis} showed 
that very few correct patches are in the search spaces of state-of-the-art techniques, which puts an upper limit on 
the number of correct patches that a G\&V technique can generate.  
It is possible to extend the search space of \gv techniques. However,
a previous study~\cite{long2016analysis} showed that this solution
failed to help existing approaches find more correct patches. It can
even reduce the number of correct fixes that these techniques can produce.
Therefore, the G\&V field is in need of a novel approach that can generate
and search a large search space in a more intelligent and scalable manner.

Neural machine translation 
is a popular deep-learning (DL) approach that
generates likely sequences of tokens given an input sequence. 
NMT has mainly been applied to natural language translation tasks
(e.g., translating French to English). In this case, the model captures
the semantic meaning of a French sentence and produces a semantically 
equivalent sentence in English.

Recently, NMT models have been used for program synthesis~\cite{alexandru2016guided},
where a model is trained to ``translate'' specifications in English to source
code. Other studies have been conducted to 
detect and repair small syntax~\cite{santos2017finding} and 
compilation~\cite{gupta2017deepfix} issues (e.g., missing parenthesis). 
While these models show promising results for fixing 
compilation issues, they only learn the syntax of the programming language 
(i.e., although the fixed program compiles and are syntactically correct, 
they often still contain runtime bugs). 
Additionally, DeepFix~\cite{gupta2017deepfix} 
showed the limitations of such approaches when repairing statements containing more than 15 characters.

Existing work applying NMT to repair bugs generally uses the best model with the ``best parameters'' based on a validation set.
Due to the diversity of bugs and fixes, such approaches struggle to cope with the complexity of program repair.
The current state-of-the-art NMT approach, SequenceR~\cite{chen2018sequencer}, performs significantly worse on the Defects4J benchmark than most
~\gv techniques that use handcrafted rules to generate repairs.
In addition, NMT-based program repair approaches struggle with representing the context of a bug.  Although context is important for fixing bugs, 
there is not yet an effective approach for incorporating context to the DL models. Naively adding the bug context to the input of the models
makes the input sequences very long. 
As a result, the models struggle to correctly capture meaningful information and such approaches can only fix short methods. For example, Tufano et al.~\cite{chen2018sequencer}
focus on methods that contain fewer than 50 tokens.
Further more, it is challenging to obtain meaningful context for millions of training instances since such training instances are partial code snippets that are not always compilable.
As a result, previous work used a reduced training set of 35,578 instances~\cite{chen2018sequencer} while most NMT techniques in other domains need millions of training instances to perform 
well. For example, training sets for translation tasks contain between 2.8M and 35.5M pairs of sentences depending on the language~\cite{gehring2017convolutional}.

\begin{figure}
\begin{lstlisting}[language=Java, basicstyle=\ttfamily\footnotesize, breaklines,escapechar=|,linebackgroundcolor={
	\ifnum \value{lstnumber}=5
	\color{red!10}
	\fi
	\ifnum \value{lstnumber}=6
	\color{green!10}
	\fi
	}]
1 import java.util.*;
2 public static int max_sublist_sum(int[] arr) {
3  int max_ending_here, max_so_far = 0;
4  for (int x : arr) {
5 - max_ending_here=max_ending_here+x;
5 + max_ending_here=Math.max(0,max_ending_here+x);
6   max_so_far = Math.max(max_so_far,max_ending_here);
7  }
7  return max_so_far;}}
\end{lstlisting}
\caption{A QuixBugs program fixed by {\tool}}
\label{motivation}
\end{figure}

In this paper, we propose a new G\&V technique called~\tool that leverages an \emph{ensemble} NMT architecture to generate patches.
This new architecture consists of an ensemble of convolutional NMT models that have different levels of complexity  
and capture different information about the repair operations. Combining them allows ~\tool to learn 
different repair strategies that are used to fix different types of bugs.
By ignoring the bug context while relying on a large training set, we significantly reduce the size of the input sequences, 
allowing us to fix bugs independently from the size of their context. 

G\&V methods are often based on complex hard-coded rules that require advanced domain knowledge and are programming language dependent.
On the other hand, an NMT-based approach learns the rules directly from the data and can be trained to fix bugs in different programming languages without a major redesign of the technique. 
Thus, ~\tool can fix bugs that are not covered by such rules and is easily applicable to other programming languages.

~\tool is trained on up to millions of pairs of buggy and fixed lines 
and produces patches that are validated against the program's test suite. 
~\tool correctly fixes the bug in Figure~\ref{motivation}.
Using the buggy line as input (line 5, in red), \tool generates a correct patch (line 5, in green) that is identical to the developer's fix. 

This paper makes the following contributions:

\begin{itemize}[leftmargin=*]
\item  A new NMT model that uses an ensemble approach to capture the diversity of bug fixes.
We show that this model performs better than the standard NMT models for automatic program repair (APR).

\item A new end-to-end automatic program repair technique, ~\tool, that leverages NMT to generate bug fixes automatically and
validate generated patches against the test suite. 
When evaluated on two popular Java benchmarks (Defects4J and QuixBugs), our technique fixes~\alljava bugs, including~\uniquejava that have not 
been fixed by any existing APR tools. 

\item An empirical study of the type of bugs that are successfully fixed by~\tool. 
We show that ~\tool can fix very diverse bugs including condition modification, variable replacement or return statement modification.

\item A study showing the portability of ~\tool to different programming languages.
 Our approach has been extended to three additional programming languages with little manual 
effort and respectively fixed \pythonfix, \cppfix, and \jsfix bugs in the Python, C++, and JavaScript benchmarks.

\item A use of attention maps to explain why certain fixes are generated or not by \tool. 

\end{itemize} 

\section{Background and terminology}
\label{back}
We present DL background necessary to understand
~\tool.

\noindent\textbf{Terminology:}
A DL {\em network} is a structure (i.e., a graph) that contains nodes or {\em layers} that are stacked
to perform a specific task.
Each type of layer represents a specific low-level transformation 
(e.g., convolution, pooling) of the input data with specific parameters (i.e., {\em weights}).
We call DL {\em architecture} an abstraction of a set of DL 
networks that have the same types and order of layers but do not specify the 
number and dimension of layers. 
A set of {\em hyper-parameters} specifies how one consolidates an architecture to a network (e.g., it defines
the convolution layer dimensions or the number of convolutions in the layer group). 
The hyper-parameters also determine which optimizer is used in training along with the optimization parameters, such as the learning rate and momentum. 
We call a {\em model} (or trained model), a network that has been trained, which has fixed weights.

\noindent\textbf{Attention:}
The attention mechanism~\cite{bahdanau2014neural} is a recent DL improvement.
It helps a neural network to focus on the most important features.
Traditionally, only the latest hidden state of the encoder are fed to the decoder.
If the input sequence is too long, some information regarding the early tokens
are lost, even when using LSTM nodes~\cite{bahdanau2014neural}.
The attention mechanism overcomes this issue by storing these long-distance dependencies in a separate attention map and feeding them to the decoder at each time step.

\noindent\textbf{Neural Machine Translation (NMT):}
NMT models typically leverage the Recurrent Neural Network (RNN) architecture to
generate the most likely sequence of tokens given an input sequence.
Long short-term memory (LSTM)~\cite{hochreiter1997long} and 
Gated Recurrent Unit~\cite{cho2014learning} are the two most popular
RNN layers for NMT.
A popular NMT approach is the encoder-decoder method~\cite{cho2014learning},
also called sequence-to-sequence (seq2seq). 
The encoder, consisting of a stack of layers, processes a sequence of tokens of 
variable length (in our case, a buggy code snippet)
and represents it as a fixed length encoding. The decoder translates this representation
to the target sequence (in our case, a fixed code snippet).
\section{Approach}
\label{sec:approach}

\begin{figure*}[tb]
	\centering
	\includegraphics[width=\textwidth]{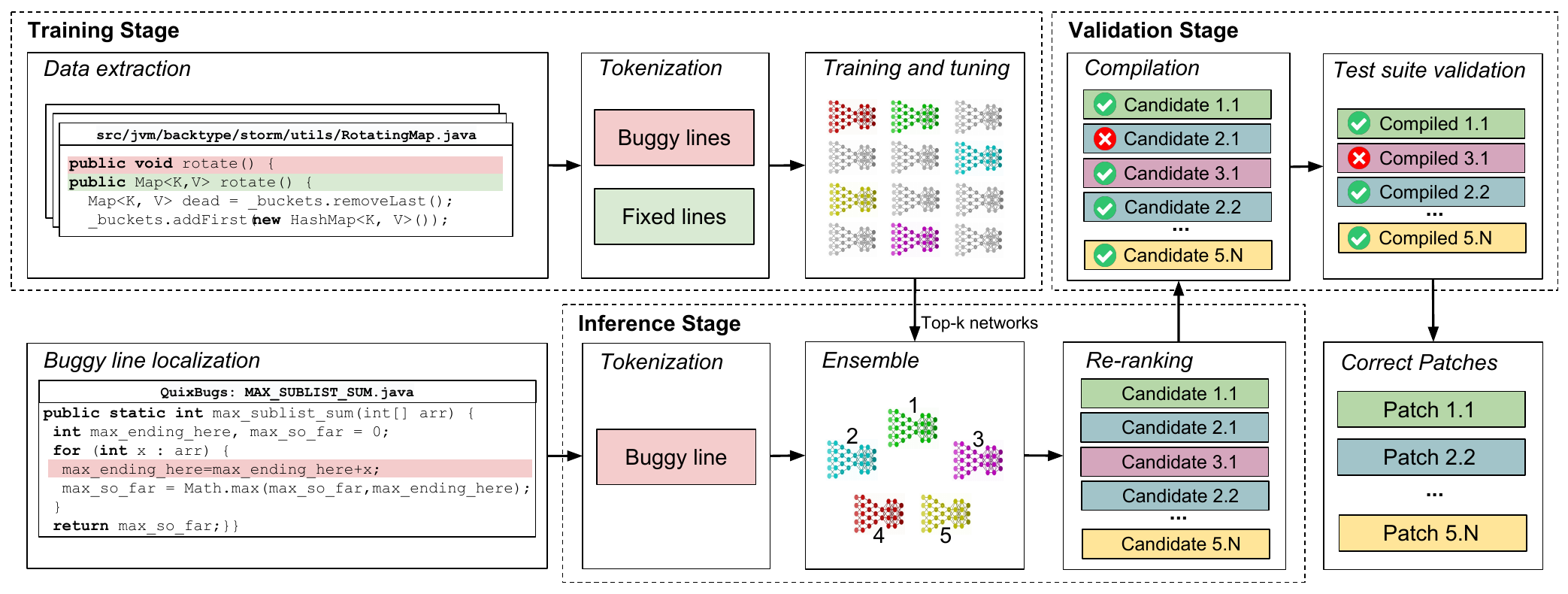}
	\caption{{\tool} overview}
	\label{fig:approach}
\end{figure*}

Our technique contains three stages: training, inference, and validation.
Figure~\ref{fig:approach} shows an overview of \tool. 
In the training phase, we extract pairs of buggy and fixed lines from 
open-source projects. Then, we preprocess these lines to obtain sequences of tokens 
and feed the sequences to an NMT network that is tuned with different sets of hyper-parameters for one epoch (i.e., a single pass on the training instances).
We further train the top-k models until convergence to obtain an ensemble of k models.
Since each of the k models has different hyper-parameters, each model learns different information that
helps fix different bugs.

In the inference phase, a user inputs a buggy line
 into ~\tool. It then tokenizes the input and feeds it to the top-k best models,
 which each outputs a list of patches. 
 These patches are then ranked and validated 
 by compiling the patched project. If a test suite is available, ~\tool runs 
 it on the compilable fixes to further filter incorrect patches.
The final output is a list of {\em candidate patches} that pass the validation stage.

Section~\ref{sec:challenges} presents the challenges of using NMT to automatically fix bugs, while the 
rest of Section~\ref{sec:approach} describe the different components of \tool.

\subsection{Challenges}
\label{sec:challenges}

APR presents several challenges, in particular when using NMT:

\noindent\textbf{(1) Choice of NMT architecture:} 
Due to the complexity of the task and the diversity of transformations used to fix bugs,
traditional NMT architectures such as LSTM~\cite{luong2015effective}, do not perform well for APR. Previous work addressed this issue by limiting the scope of
the repair to simple compilation errors~\cite{gupta2017deepfix} or by focusing on a few specific transformations~\cite{devlin2017semantic}.
However, these solutions are not ideal since it restricts the model to fixing a limited range of bugs.
We address this issue by using a new type of NMT architecture, called fconv~\cite{gehring2017convolutional}, that relies on convolutional layers instead of RNN layers. 
This architecture performs better because the convolutional layers better capture immediate context information than RNN layers while the multi-step attention
allows the architecture to keep track of long-term dependencies. This type of architecture has been shown to work well for grammatical error correction~\cite{Chollampatt2018}   
and translation~\cite{gehring2017convolutional}
but has not been applied to other domains.

\noindent\textbf{(2) Diversity of Bug Repair:}
Bugs are very diverse, with different fixes needed to fix the same buggy line depending on the context.
Since an NMT model has no access to the entire project (due to limitation of sequence size of existing NMT models), the model might not have enough
information to generate a correct patch 
for its first try.
In addition, a single model might overfit the training data and fail to capture the diverse fix patterns.
To address this challenge, we propose an ensemble approach that combines models with different hyper-parameters,  
to increase the diversity of patches generated, combined with a validation stage that is used to discard incorrect patches. Our ensemble approach allows us to 
fix 47\% more bugs than using a single model.

\noindent\textbf{(3) Adaptability to different Programming Languages:}
Existing APR techniques rely on handcrafted fix patterns that require domain knowledge to create and are not easily transferable to
different languages.
Leveraging NMT allows \tool to learn how to fix bugs instead of relying on handcrafted patterns. Thus, 
\tool is generally applicable to other programming language, as we can obtain training data automatically. 
\tool{} is the first approach to successfully fix bugs in four popular programming languages.
 
\noindent\textbf{(4) Large vocabulary size:} 
Compared to traditional natural language processing (NLP) tasks such as translation, the vocabulary size of 
source code is larger and many tokens are infrequent or composed of multiple words.
In addition, letter case indicates important meanings in source code (e.g., \texttt{Zone} is generally a class, \texttt{zone} a variable, 
and \texttt{ZONE} a constant), which increases further the vocabulary size. For example, previous work~\cite{chen2018sequencer} had to 
handle a code vocabulary size larger than 560,000 tokens. Since such a high number of token is not scalable for NMT, practionners 
need to cut the vocabulary size significantly, which leaves a large number of infrequent tokens out of the vocabulary.
We address this challenge by using a new tokenization approach that reduces the vocabulary size significantly without
increasing the number of words out of the vocabulary. Overall, less than 2\% of tokens in our test sets are out of the vocabulary (Section~\ref{sec:tok}).

\subsection{Data Extraction}
We train ~\tool on pairs of buggy and fixed lines of code
extracted from the commit history
of 1,000 open-source projects. To remove commits that are not related to bug fixes,
 we apply a few filters. First, we  only keep changes that have the words 
 ``fix,'' ``bug,'' or ``patch'' in their associated commit message. This is
 a standard method that has been done in previous work~\cite{wang2016automatically}. 
 However, by manually investigating a random sample of 100 pairs,
 we found that these patterns were not enough to filter unrelated commits.
 Indeed, only 62/100 pairs were related to real bug fixes. The main reason
 was that many developers use the keywords ``fix'' and 
 ``patch'' in commits that are unrelated to bugs.
To address this issue, we also exclude commits that contain the 
following six anti-patterns in their messages: ``rename,'' 
``clean up,'' ``refactor,'' ``merge,'' ``misspelling,'' and ``compiler warning.''
Using these anti-patterns increases the number of correct pairs in our 
random sample to 93.

We further remove sequences which size is than 2 times the standard
deviation, cosmetic changes (e.g., spacing and indentation).
We also removes comments and sequences that contain non-utf-8 characters.
We focus on single-statement changes as smaller sequences 
are easier to learn for NMT models. 
Finally, to propose a fair comparison with other APR techniques, we remove from our training data 
all changes that are identical to bug fixes in the Defects4J benchmark.
After filtering, we obtain~\trainingdata pairs of buggy and fixed lines.

\subsection{Input Representation}
\label{sec:tok}

\textbf{Input Sequences:}
The input of ~\tool is a tokenized buggy line. We chose this simpler representation
over representations used in previous work (e.g. full function as input) for the
following reasons:

\noindent
\textbf{(1)} NMT approaches work well with small sequences, but struggle when an input sequence contains more than 30 tokens~\cite{cho2014properties}. 
Adding context to the input significantly increases the size of the input sequences. For example, the median
size of previous work~\cite{chen2018sequencer} training instances is above 300 tokens. Only feeding the 
buggy line allows us to reduce the size of training instances to an average of \tokenjava tokens per sequence, making
the learning process easier.

\noindent
\textbf{(2)} Context generally contains many tokens that are unnecessary for fixing the bug. Thus, increasing the size
of the context also has the disadvantage of significantly increasing the vocabulary size. For example,
while having only 35,578 training instances, previous work's vocabulary contains 567,304 tokens.
Not using context gives us an unfiltered vocabulary of over 200,000 tokens for about 1 million instances. 
The tokenization process described below further reduces the vocabulary
to 64,044 tokens
while keeping the ratio of out-of-vocabulary tokens during inference below 2\%.

\noindent
\textbf{(3)} Although context is important for fixing bugs, there is not yet an effective
approach for incorporating large context to NMT models. In practice,
we found that straightforward approaches of adding context do not increase
the number of bugs fixed.

\noindent\textbf{Tokenization:}
Typical NMT approaches take a vector of tokens as input. Therefore, our first challenge is to choose
a correct abstraction to transform source code into sequences of tokens.

The lowest level of tokenization we can use is character-level tokenization. 
This has the advantage of using a small set of different tokens. However, this level of
tokenization makes the training of the model harder as it has to learn what
a ``word'' is before learning correct statements. This tokenization method
has been used successfully to fix compilation errors in DeepFix but showed
limits for statements that contained more than 15 characters~\cite{gupta2017deepfix}.

For most natural languages, using word-level tokenization provides better results
than character-level tokenization~\cite{mikolov2012subword} so we decided
to use word-level tokenization. 
While source-code is analogous to natural language, word-level tokenization 
presents several challenges that are specific to programming languages.

First, the vocabulary size becomes extremely large and many words are infrequent 
or composed of multiple words without separation (e.g., getNumber and get\_Number are two different words). To address this 
issue specific to source-code generation, we enhanced the word-level 
tokenization by also considering underscores, camel letter and numbers
as separators.
Because we need to correctly regenerate source code from the list of tokens generated by the NMT model, we also need to
introduce a new token (\texttt{<CAMEL>}) :wqfor camel cases, in order to mark where the split 
occurs. 
In addition, we abstract all strings in one specific token
``STRING'' and all infrequent numbers  (i.e., different from 1 and 0) to ``NUMBER.''
The reason for the string abstraction is that tokens in strings represent a completely different language that might confuse the model.

Previous work~\cite{tufano2018empirical} uses a complete abstraction of all tokens of the buggy line. 
We do not do it for several reasons: first, such abstraction loses the semantics contained in the variable name. We believe that such semantics provide
useful information to the network. Second, the abstraction is implicitly included in the original code snippet. If the network needs it to repair a bug,
the model will learn it from the source code. In fact, the provided abstraction might not be the best abstraction for the network. 

\subsection{NMT Architecture}
\begin{figure}[tb]
	\centering
	\includegraphics[width=0.49\textwidth]{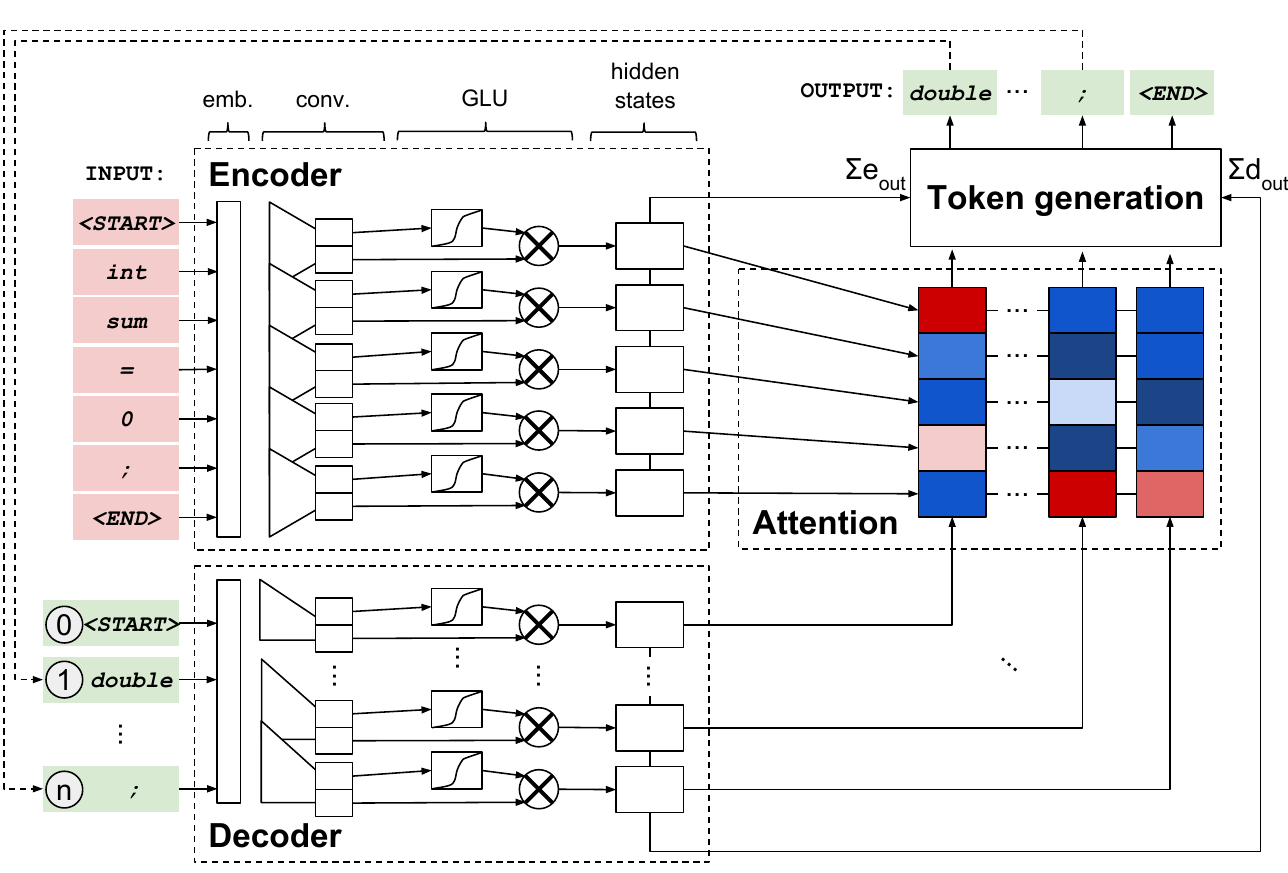}
	\caption{The NMT architecture used in ~\tool.}
	\label{fig:net}
\end{figure}

Figure~\ref{fig:net} shows an overview of \tool's neural machine translation architecture. 
For simplicity, Figure~\ref{fig:net} only displays a network with one convolution layer.
In practice, depending on the hyper-parameters, a complete network has 2 to 10 convolution layers for each encoder and decoder.
Our architecture consists of three main components: an encoder, a decoder,
and an attention module.

In training mode, the model has access to both the buggy and the fixed lines.
The model is trained to generate the best representations of the transformation from
buggy to fixed lines. In practice, this is conducted by finding the best combination
of weights that translates buggy lines in the training set to fixed lines.
Multiple passes on the training data are necessary to obtain the best set
of weights. 

In inference mode, since the model does not have access to the fixed line, 
the decoder processes tokens one by one, starting with a generic <START> token. 
The output of the decoder and the encoder are then combined through
the multi-step attention module. Finally, new tokens are generated based on the
output of the attention, the encoder and the decoder.
The generated token is then fed back to the decoder until the <END> token is generated. 

Following the example input in Figure~\ref{fig:net},
a user inputs the buggy statement ``\texttt{int sum=0;}'' to \tool.
After tokenization, this sequence is fed to the encoder which 
generates an encoded representation ($e_{out}$).
Since \tool did not generate any token yet, the token generation starts by 
feeding the token <START> to the decoder (iteration 0) 
The output of the decoder ($d_{out}$) is then 
combined with the encoder output using a dot product to form the first column
of the attention map. The colors of the attention map indicate how important
each input token is for generating the output. For example, to
generate the first token, the tokens \texttt{int} and \texttt{0} are the 
two most important input tokens as they appear in dark and light red respectively.
 The token generation combines the output of the attention,
as well as the sum of the encoder and decoder outputs ($\Sigma$$e_{out}$ and $\Sigma$$d_{out}$)
to generate the token \texttt{double}. This new token is added to the list 
of generated tokens and the list is given back as a new input to the decoder (iteration 1).
The decoder uses this new input to compute the next $d_{out}$ that is used to 
build the second column of the attention map and generate the next token.
The token generation continues until the token <END> is generated.
 
We describe below the different modules of
the network.

\noindent\textbf{Encoder and Decoder:}
The purpose of the encoder is to provide a fixed length vectorized representation of the input sequence 
while the decoder translates such representation to the target sequence (i.e., the patched line).
Both modules have a similar structure that consists of three main blocks: an embedding layer,
several convolutional layers, and a layer of gated linear units (GLU).

The embedding layers represent input and target tokens as vectors, with tokens occurring in similar 
contexts having a similar vector representation. In a sense, the embedding layers represent the 
model's knowledge of the programming language. 

The output of the embedding layer is then fed to several convolutional layers. The size of the convolution 
kernel represents the number of surrounding tokens that are taken into consideration. 
Stacking such convolutional layers provides multiple levels of abstraction for our network to work with.
The convolution layers also provide information regarding surrounding tokens. The number
of tokens considered depends on the size of the convolution kernels. This layer is different
in the encoder and the decoder.
The encoder uses information from both the previous and the next tokens in the input sequence (since the full input sequence is known at all time),
while the decoder only uses information about the previously generated tokens (since the next tokens have not been generated yet).
This difference is represented in Figure~\ref{fig:net} by full triangles in the encoder and half triangles in the decoder.

After the convolutional layers, a layer of GLU (represented by the sigmoid and multiplication boxes in Figure~\ref{fig:net}). 
is used to decide which information should be kept by the network. 
For more details about the encoder and decoder, refers to previous work~\cite{gehring2017convolutional}.

\noindent\textbf{Multi-Step Attention:}
As mentioned in Section~\ref{back}, the attention mechanism helps to keep track of early information
by connecting all the hidden states of the encoder to the decoder instead of only the last hidden state.
In practice, this is done with a simple dot product between the output of the 
encoder and decoder convolution layers.

Compared to traditional attention, multi-step attention uses an attention
mechanism to connect the output of each convolutional layers in the encoders and decoder.
When multiple convolutional layers are used, it results in multiple attention maps.
Multi-step attention is useful because it connects each level of abstraction (i.e., convolutional layers) to the outputs. 
This increases the amount of information from the decoder that is shared with the decoder when generating the target tokens.

The attention map represent the impact of input tokens for generating a specific token and can help understand
why a specific output is generated. As shown in Figure~\ref{fig:net},
the rows of the map represent the input while the columns represent the generated tokens. For each generated token, 
the attention map shows which input tokens are the most important. For example, in Figure~\ref{fig:net}, 
the most important tokens to generate \texttt{double} were the first one (\texttt{int}) and the second last one (\texttt{0}).
We analyze attention maps to answer RQ5.

\noindent\textbf{Token Generation:}
The token generation combines the output of the attention layers, encoder ($\Sigma$ $e_{out}$) and decoder ($\Sigma$ $d_{out}$)
to generate the next token. Each token in the vocabulary is ranked by the token generation component based
on their likelihood of being the next token in the output sequence. If we are only interested in the top-1 result generated by the model,
the most likely token is selected and appended to the list of generated tokens. The list is then
sent back as the new input of the decoder. The token generation stops when the <END> token is generated.

\noindent\textbf{Beam Search:}
Since our model has no information about the context of the buggy line, it is unlikely for the first patch generated by the model to be correct.
Therefore, we want \tool to generate multiple patches that will then be validated.
Generating and ranking a large number of patches is challenging because since the patches are generated token by token, 
we cannot know the probability of the final sequence before generating all the tokens. 
and choosing the most likely token at each iteration might not lead to the most likely sequence.
To address this issue, we use a search strategy called beam search that is commonly used for NMT.
The goal of beam search is to find the most likely sequence instead of the most likely token at each step. 
For each iteration, the beam search algorithm checks the t most likely tokens (t corresponds to the beam width)
and ranks them by the total likelihood score of the next s prediction steps (s correspond to the search depth).
In the end, the beam search algorithm outputs the top t most likely sequences ordered based on the likelihood of each sequence.

\subsection{Ensemble Learning:}
\label{sec:ensemble}

Fixing bugs is a complex task because there are very diverse bugs with very different fixing patterns that
vary in term of complexity.
Some fix patterns are very simple (e.g., to change the operator $\textless{}$ to $\textgreater$)
while others require more complex modifications (e.g., adding a null checker or calling a different function).
Training an effective generalized model to fix all types of bug is difficult. 
Instead, it is easier to overfit models to fix specific types of bugs. Then, we can combine these models
into a general ensemble model that will fix more bugs than one single model. 

Therefore, we propose an ensemble approach that combines models with different hyper-parameters (different networks)
that performed the best on our validation set.
The complexity of the model is represented by hyper-parameters such as the number and dimension of
convolutional layers in the encoder and decoder.

As described in Section~\ref{back}, hyper-parameters are parameters that consolidate an architecture to a network.
Hyper-parameters include the number of layers, the dimensions of each layer and specific rates
such as the learning rate, the dropout or the momentum. Different hyper-parameters have a large
impact on the complexity of a network, the speed of the training process and the final performance of the trained model.
For this tuning process, we chose to apply random search because
previous work showed that it is an unexpensive method that performs better than other common hyper-parameter tuning strategies such as grid search and manual search~\cite{bergstra2012random}.
For each hyper-parameter, we define a range from which we can pick a random value.
Since training a model until convergence is very expensive, the tuning process is generally for
only one epoch (i.e., one pass on the training data). 
We trained $n$ different models with different random sets of parameters to 
obtain models with different behavior and kept the top $k$ best model based on the performance of each model
on a separate validation set.

One challenge of ensemble learning is to combine and rank the output of the
different models. In our case, the commonly used majority voting would not work well
since we specifically chose models that are likely to generate different 
fixes. Instead, we use the likelihood of each sequence (i.e., fixes) generated by each model.
Since we use beam search to generate the top-k patches for each model, each 
patch has an associated negative log likelihood. We use this score to rank the output of 
all models. When two models produce the same patches with different 
scores, we only consider the highest score. The intuition for selecting the
higher score instead of the average is that the models are designed to fix
different types of bugs, so if one model is confident that one specific patch is correct, its confidence should not
be impacted by other models.

\subsection{Patch Validation}

\noindent\textbf{Statement Reconstruction:}
Our model outputs a list of tokens that form a fix for the input buggy line.
The statement reconstruction module
generates a complete patch from the list of tokens. This step is mostly an 
inversion of the tokenization step. However, there are two abstractions that
cannot be reversed to source code directly: the \textless STRING\textgreater~ 
and \textless NUMBER\textgreater~ tokens.
For these two tokens, we extract candidate numbers and strings from the original
buggy line and attempt to replace the corresponding tokens with them. 
If there are no strings or numbers in the buggy line,
the patch is discarded.
Since there are not many different numbers or strings in one line of code, the impact
of the total number of reconstructed patches is negligible.
Once the fix is generated, it is inserted at the 
buggy location and we move to the validation step.

\noindent\textbf{Compilation and Test Suite Validation:}
The neural network does not have access to the entire project; therefore
it does not know whether the generated patches are compilable or pass the test 
suite.
For this reason, we use a validation step to filter out patches that do not compile or do not pass the
triggering test cases. This step is similar to the validation process
done by traditional APR approaches with one difference. 
We do not require all the test
cases to pass to consider a patch plausible. 
Indeed, there can be multiple
bugs in one projects and several test cases might fail because of another bug~\cite{yang2017better}.
Therefore, even the fixed version might still fail for some test cases.
To alleviate this issue, we use the same two criteria as previous work~\cite{yang2017better}.
First, the test cases that make the buggy version pass should still pass on the patched version.
Second, test cases that failed on the version fixed by developers can still fail.

\subsection{Generalization to Other Languages}
Since \tool learns patterns automatically instead of relying on handcrafted
patterns, it can be generalized to other programming languages with minimum
effort. The main change required is to obtain new input data in the correct language.
Fortunately this is easy to do since our training set is extracted from open-source
GitHub projects. Once the data for the new programming
language has been extracted, the top k models can be retrained without any re-implementation. 
\tool will learn fix patterns automatically for the new programming language.
\section{Experimental Setup}
\label{sec:setup}
\begin{table}[t]
\centering
\caption{Training dataset information. \# token PL indicates the average number of tokens per line of code.}	
\label{tab:dataset_info}
\resizebox{0.99\columnwidth}{!}{
\begin{tabular}{@{}lrrrrrr@{}}
	\toprule
	\textbf{Language} & \textbf{\# projects} & \textbf{\# bug fixing commits} &\textbf{\# instances}    & \textbf{src. Vocabulary size} & \textbf{trg. Vocabulary size} &  \textbf{\# token PL} \\
	\midrule
	Java       & 1,000 & 1,752,212                                            & \trainingdata           & 51,703  & 49,303  &   \tokenjava \\
	Python     & 1,000 & 1,717,249                                            & \trainingdatapython     & 78,999  & 61,687  & \tokenpython \\
	C++        & 1,000 & 4,466,504                                            & \trainingdatacpp        & 234,039 & 222,583 & \tokencpp \\
	JavaScript & 1,000 & 1,341,121                                            & \trainingdatajavascript & 43,919  & 38,543  & \tokenjavascript \\
	\bottomrule
\end{tabular}
}

\end{table}

\vspace{2mm}
\noindent
\textbf{Dataset:}
To train our technique on different programming languages,
we collect data from the top 1,000 GitHub projects (based on star rating) in 4 popular languages
(i.e., Java, Python, C++, and JavaScript).
Table~\ref{tab:dataset_info} presents a summary of our dataset.
We extract more training samples for Java and C++ because these two languages contain many extremely large projects
such as Linux and Apache projects.
To tune hyper-parameters, we pick a random sample of 2,000 instances as our validation dataset and use the rest for training.

We evaluate \tool on five benchmarks.
For Java, we use Defects4J~\cite{just2014defects4j} and QuixBugs~\cite{lin2017quixbugs}. 
For Python, we use Python's version of QuixBugs. For C++, we used the 69 real-world defects from prior work~\cite{long2016automatic,long2015staged}. 
Since there is no automatic repair benchmark in JavaScript, we use the 12 examples associated with common bug patterns in JavaScript described in prior work~\cite{hanam2016discovering}.

To ensure fairness, if a bug appears in both our test and training sets (or both our test and validation sets), we remove the bug 
from our training set, which is rare anyways. 

\vspace{2mm}
\noindent
\textbf{Training and Tuning:}
We use random search to tune hyper-parameters.
We limit the search space to reasonable values: embedding size (50-500), convolution layer dimensions (128*(1-5), (1-10)), 
number of convolution layers (1-10), drop out rate (0-1), gradient clipping level (0-1), learning rate (0-1), and momentum (0-1).
We first uniformly pick a random set of hyper-parameters within the search space. Then, we train the model for one epoch using the training data. 
The trained model is evaluated using the validation set. We repeat the process for five days and rank the hyper-parameters sets based on their 
perplexity measures~\cite{jelinek1977perplexity}, which is a standard metrics in NLP that measure how well
a model generates a token.

\vspace{1mm}
\noindent
\textbf{Infrastructure:}
We use the implementations of LSTM, Transformer, and FConv provided by fairseq-py~\cite{fairseqpy}
running on Pytorch~\cite{paszke2017pytorch}.
Our models were trained and evaluated on an Intel Xeon E5-2695 and two Gold SKL 5120 machines 
and NVIDIA TITAN V and Xp GPUs.

\vspace{2mm}
\noindent
\textbf{Performance:}
The median time to train our NMT model for 1 epoch during tuning is 38 minutes.
Training our top 5 models sequentially until convergence took 93 hours (23 hours if training concurrently on the two Xeon servers).
In inference, generating 1000 patches for a bug takes on average 8 seconds. 
\section{Evaluation and Results}
\label{eval}
\tool generates
a list of candidate patches that successfully pass all the bug triggering
test cases. 
For evaluation purpose only, we manually compare the candidate patches to 
the developer patch and consider a patch as a correct fix if it is identical or semantically equivalent to the developer patch. 
Upon acceptance, we will make available the hyper-parameter values of our ensemble models, the list of bugs correctly fixed by \tool, 
the top 5 trained models, as well as the patches generated by \tool.

\subsection{RQ1: How does {\tool} perform against state-of-the-art NMT models?}
\noindent\textbf{Approach:}
We compare ~\tool with SequenceR~\cite{chen2018sequencer}, the state-of-the-art program NMT-based APR and 
two other state-of-the-art NMT architectures (i.e., LSTM~\cite{luong2015effective} and Transformer~\cite{vaswani2017attention}).
These two models have not been used for program repair so we implemented in the same framework as our work (i.e., using Pytorch~\cite{paszke2017pytorch} 
and the fairseq-py~\cite{fairseqpy}  
library). We then trained and tuned them similarly to ~\tool.
SequenceR~\cite{chen2018sequencer} is an approach on arXiv concurrent to \tool that uses NMT to automatically repair bugs. We use the numbers reported by SequenceR's authors 
on the Defects4J dataset for comparison.
We cannot compare with SequenceR on the QuixBugs dataset because they did not run their tool on the QuixBugs dataset
and the tool is unavailable. 

Following SequenceR~\cite{chen2018sequencer}
we assume perfect localization to ensure a fair comparison. 
As mentioned in previous work~\cite{liu2018you}, the effect of fault localization on automatic repair techniques should be analyzed separately, since different localization techniques 
provide significantly different results~\cite{liu2018you}.

\noindent\textbf{Comparison with State-of-the-art NMT:}
\tool with $k$=$5$ fixes \djfix and \quixfix bugs in the Defects4J and QuixBugs benchmarks respectively, including ~\djfirst and \quixfirst that are ranked first.
Specifically, Table~\ref{tab:comp_dl_approaches} displays the number of bugs fixed by the different approaches on our two benchmarks.
The Fconv column shows the results for our approach without ensemble.
In the top-1 column, we report the number of bugs that were fixed correctly with the first candidate patch generated by the tool. The column ``All'' display the total
number of bugs each technique can fix, regardless of the ranking.
Our technique fixes the most number of bugs considering the top-1 candidate, with \djfirst  bugs fixed in the Defects4J benchmark  and \quixfirst in the QuixBugs benchmark.
Considering all correct patches, 
~\tool also fixes the most bugs with \djfix correct fixes in the Defects4J benchmark and \quixfix  in the QuixBugs benchmark. 

\tool fixes all bugs fixed by the LSTM and Transformer approaches and most of the bugs fixed by SequenceR. 

\begin{figure}[tb]
	\centering
	\includegraphics[trim={10pt 0 0 0},clip,width=0.47\textwidth]{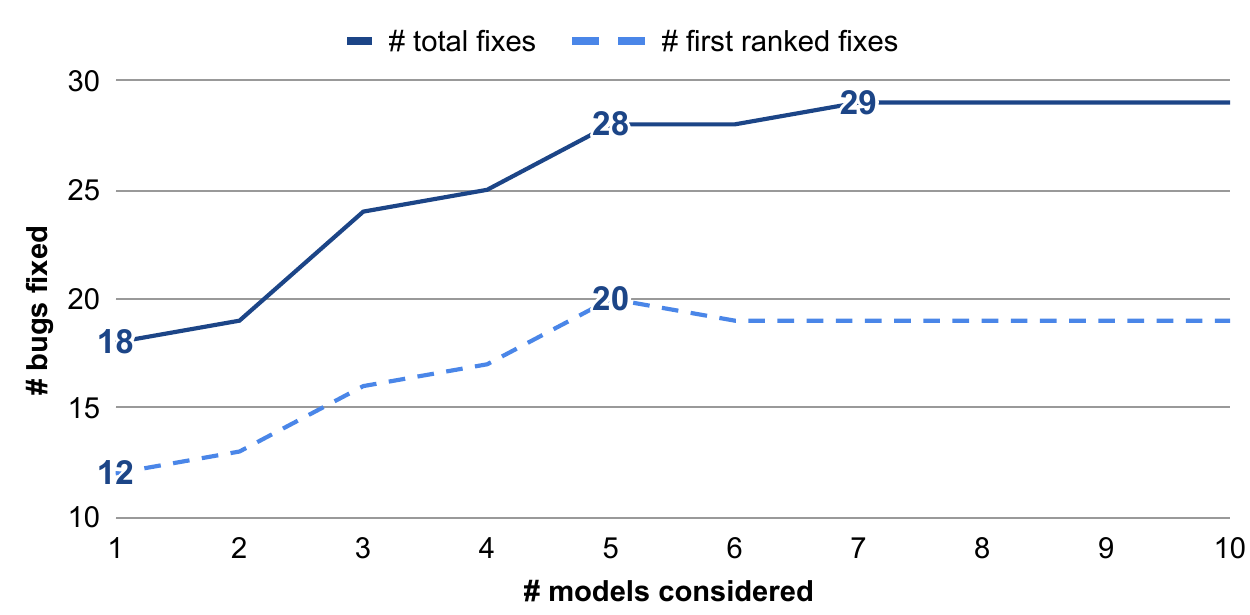}
	\caption{Number of bugs fixed as the number of models considered in ensemble learning increases.}
	\label{fig:ens}
\end{figure}

\noindent\textbf{Advantage of Ensemble Learning:}
To demonstrate the advantage of our ensemble learning approach, Figure~\ref{fig:ens} shows the total number of bugs fixed using the top-k models, with $k$ varying from 1 to 10.
With $k$=$1$ (i.e., without using ensemble), our model only fixes 18 bugs in the Defects4J dataset. As $k$ 
increases, the number of bugs fixed increases, until reaching a plateau of 29 bugs for $k$=$7$.

Since increasing $k$, increases the number of models considered, the number of generated patches increases too. 
Thus, it might have a negative impact on the ranking of correct fixes.
The light blue dotted line in Figure~\ref{fig:ens} displays the evolution of the number of patches that are ranked first when the number of models increases.
Surprisingly, this evolution is very similar to the evolution of the total number of bugs fixed and adding more models does not significantly reduced the number
of correct patches ranked first.

Ensemble learning provides a significant improvement compared to using one single model. With $k$=$5$, the ensemble model fixes 47\% more bugs than with a single model, with 66\% more correct patches
ranked first.

\noindent\textbf{Impact of Hyper-parameter tuning:}
To ensure our results are the consequence of using different hyper-parameters and not because of the randomness in training, we train our first model 8 times with identical hyper-parameters.
The average number of bugs
fixed in the Defects4J benchmark by this first model is 18.9, with a variance of 0.125. This shows that using an ensemble approach of the exact same model would not work well since the
randomness in training does not have too much impact on the evaluation results.

\begin{table}[t]
\centering
\caption{Comparison with other deep learning approaches.}
\label{tab:comp_dl_approaches}
\adjustbox{max width=0.99\columnwidth}{

\begin{tabular}{@{}llrrrrr@{}}
\toprule
&       & \multicolumn{2}{c}{\textbf{Baselines}} & \multirow{2}{*}{\textbf{SeqR}~\cite{chen2018sequencer}} & \multirow{2}{*}{\textbf{Fconv}} & \multirow{2}{*}{\textbf{\tool} (k=5)} \\
\cmidrule{3-4}
&       & \textbf{LSTM} & \textbf{Trans.} &  &  & \\
\midrule
\multirow{2}{*}{\textbf{Defects4J}} & Top-1 &   6         & 6                 & 12             & 12 & \textbf{ \djfirst}  \\  
                                    & All   &  10         & 9                 & 14             & 18 & \textbf{ \djfix}  \\ 
\midrule
\multirow{2}{*}{\textbf{QuixBugs}}  & Top-1 &   8         & 9                 & NA             &  6  & \textbf{  \quixfirst}       \\ 
                                    & All   &   9         & 9                 & NA             &  8  & \textbf{ \quixfix}  \\
\bottomrule
\end{tabular}
}
\end{table}

\noindent\fbox{
\centering
\parbox{0.95\linewidth}{
\textbf{Summary:} Compared to state-of-the-art NMT-based approaches, \tool fixes the most number of bugs, \textbf{\alljava bugs}, across two popular Java benchmarks, 
including \textbf{\allfirstjava} that are ranked first. 
}
}

\subsection{RQ2: How does {\tool} perform against state-of-the-art APR techniques?}

\noindent\textbf{Approach:}
We compare ~\tool with state-of-the-art \gv approaches on all six projects in the Defects4J benchmark and the QuixBugs benchmark.
We extracted the results for each technique from previous work~\cite{liu2018you} and cross-checked against the original paper of each technique.
As we did for RQ1, we only consider bug fixes that have been manually verified as correct and are ranked first by the APR technique. 
For \tool we also show in parentheses the number of bugs fixed that have not been fixed by other techniques.

\noindent\textbf{Results on the Defects4J benchmark:}
Table~\ref{tab:comp_trad_approaches} shows how \tool performs against state-of-the-art~\gv approaches.
\tool generated~\djfirst fixes that are ranked first, making it close to the best six ~\gv APR technique.
and outperforming 11 techniques.
In addition, three of the patches ranked first have not been fixed before, indicating that \tool can potentially fix different bugs than previous work.
If we consider correct fixes generated by ~\tool independently from ranking, it is the second best ~\gv approach, fixing 28 bugs, fixing 6 bugs no other techniques fixed.

\begin{table*}[t]
\centering
\caption{Comparison with state-of-the-art~\gv approaches. The number of bugs only ~\tool can fix are in parentheses.}
\label{tab:comp_trad_approaches}
\adjustbox{max width=0.99\linewidth}{
\begin{tabular}{@{}lrrrrrrrrrrrrrrrrrr@{}}
\toprule
        & \multicolumn{3}{c}{\textbf{Astor}~\cite{martinez2016astor}}  & \textbf{HDR}    &  \textbf{Nopol}      & \textbf{ACS}            & \textbf{Elixir}       & \textbf{JAID}           & \textbf{ssFix}           &\textbf{CapGen}        & \textbf{SketchFix}      & \textbf{FixMiner}          & \textbf{LSRepair}      & \textbf{SimFix}         & \textbf{SOFix}       & \textbf{SeqR}       &  \multicolumn{2}{c}{\textbf{\tool}} \\
\cmidrule(rr){2-4}\cmidrule(ll){18-19}
\textbf{Projects}         & \textbf{jGP}        & \textbf{jKali}           & \textbf{jMutR} & \cite{le2016history} & \cite{xuan2017nopol} & \cite{xiong2017precise} & \cite{saha2017elixir} & \cite{chen2017contract} & \cite{xin2017leveraging} & \cite{wen2018context} & \cite{hua2018sketchfix} & \cite{koyuncu2018fixminer} & \cite{liu2018lsrepair} & \cite{jiang2018shaping} & \cite{liu2018mining} & \cite{chen2018sequencer} & Top 1 & All \\

\cmidrule(rr){1-1} \cmidrule(rr){2-4} \cmidrule(rr){5-17} \cmidrule(ll){18-19}

Chart    & 0                        & 0                        & 1                        & 0                    & 1                    & 2                       & 4                     & 2                       & 3                        & 4                     & 6                       & 5                          & 3                      & 4                       & 5                    &  1                       &  2 (0)   &  3 (0)        \\
Closure  & 0                        & 0                        & 0                        & 0                    & 0                    & 0                       & 0                     & 5                       & 2                        & 0                     & 3                       & 5                          & 0                      & 6                       & 0                    &  3                       &  8 (2)   & 11 (5)       \\
Lang     & 0                        & 0                        & 0                        & 2                    & 3                    & 3                       & 8                     & 1                       & 5                        & 5                     & 3                       & 2                          & 8                      & 9                       & 4                    &  2                       &  4 (1)   & 4 (1)         \\
Math     & 5                        & 1                        & 2                        & 4                    & 1                    & 12                      & 12                    & 1                       & 10                       & 12                    & 7                       & 12                         & 7                      & 14                      & 13                   &  5                       &  4 (0)   & 9 (0)         \\
Mockito  & 0                  		 & 0                        & 0                        & 0                    & 0                    & 0                       & 0                     & 0                       & 0                        & 0                     & 0                       & 0                          & 1                      & 0                       & 0                    &  0                       &  0 (0)   & 0 (0)        \\
Time     & 0                        & 0                        & 0                        & 0                    & 0                    & 1                       & 2                     & 0                       & 0                        & 0                     & 0                       & 1                          & 0                      & 1                       & 1                    &  0                       &  1 (0)   & 1 (0)         \\
\cmidrule(rr){1-1} \cmidrule(rr){2-4} \cmidrule(rr){5-17} \cmidrule(ll){18-19}
\textbf{Defects4J}    & 5                        & 1                        & 3                        & 6                    & 5                    & 18                      & 26                    & 9                       & 20                       & 21                    & 19                      & 25                         & 19                     & 34                      & 23                   &  12                      &  \djfirst (3)  & \djfix (6)        \\
\cmidrule(rr){1-1} \cmidrule(rr){2-4} \cmidrule(rr){5-17} \cmidrule(ll){18-19}
\cmidrule(rr){1-1} \cmidrule(rr){2-4} \cmidrule(rr){5-17} \cmidrule(ll){18-19}
     
\textbf{QuixBugs}  &  \multicolumn{3}{c}{6~\cite{Ye2018APRQuixBugs}}                                          & NA                                    & 1                                    & NA                                   & NA                                    & NA                                    & NA                                      & NA                                   & NA                                         & NA                                           & NA                                       & NA                                      & NA      & NA                     &  \quixfirst (10)      &  \quixfix (10)        \\
\bottomrule
\end{tabular}
}
\end{table*}

\noindent\textbf{Results on the QuixBugs benchmark:}
Most \gv techniques were not evaluated on the QuixBugs benchmark, however; the ASTOR framework~\cite{martinez2016astor} 
(at the time, a combination of jKali, jGenProg,  and jMutR), as well as Nopol~\cite{xuan2017nopol}, have been evaluated on this benchmark, fixing respectively 6 and 1 bugs. 
Compared to these two approaches,
~\tool fixes \quixfix bugs (\quixfirst of them being ranked first and 10 of them being unique to \tool) significantly outperforming both techniques.

\noindent\textbf{Impact of Fault Localization:} It is difficult to conduct a fair comparison of all APR techniques since the localization results used are different. 
For example, 
HDRepair uses Ochiai~\cite{abreu2007accuracy} but assumes that the correct file and method of the bug is known. 
SimFix uses GZoltar 1.6.0 combined with test case purification, while  
approaches such as jKali and Nopol used a now outdated version of GZoltar that is less accurate.
SketchFix and ELIXIR did not report 
the localization framework they used. Such differences among the bug localization component make it difficult to do a fair comparison~\cite{liu2018you}.
Following SequenceR, we assume perfect localization. However, looking at the output of GZoltar 1.6.0 implementation of Ochiai, 
only two of the bugs \tool fixes have a missing line-level localization, indicating that for the remaining 26 of the bugs fixed 
by \tool, their localization results could be found by the state-of-the-art bug localization technique, including 4 bugs only fixed by \tool.

\noindent\fbox{
\centering
\parbox{0.95\linewidth}{
\textbf{Summary:} Considering the top-1 patches,~\tool is on par with the best ~\gv APR techniques, outperforming the rest. 
In addition, \uniquejava of the bugs fixed by \tool have not been fixed before (\uniquedj in the Defects4J benchmark and 10 in the QuixBugs benchmark).

}
}

\subsection{RQ3: What type of bugs can \tool fix?}
\noindent\textbf{Approach:} 
To understand the type of bugs fixed by \tool, we consider the list of actions 
associated with each Defects4J fixes proposed by previous work~\cite{sobreira2018dissection}. 
These actions indicate which transformations are used to repair the bugs. We also consider
which parts of the program \tool modifies to generate a correct patch.

\noindent\textbf{Results:}
~\tool performed 25 different types of action to fix bugs in the Defects4J dataset, indicating that our models learn to fix bugs in different manners.
On average, \tool performs 1.8 actions per bug fix while SequenceR performs 1.2 actions per bug fix,
indicating that \tool might be fixing more complex bugs than SequenceR.
Only 4 patches are deletions, while 11 of the bugs fixed by ~\tool require 2 or more actions to generate the correct patches, 
indicating that it can generate complex patches.
~\tool fixes bugs by modifying many different parts of a program, including 
conditional expressions (16 bugs), 
method calls (9 bugs), 
return statements (7 bugs), 
assignment (5 bugs), 
object instantiation (3 bugs), and
method definition (1 bug).
We show below some of the bugs \tool successfully fixed.

\noindent\fbox{
\centering
\parbox{0.95\linewidth}{
\textbf{Summary:} \tool fixes bugs from the Defects4J dataset by performing 25 different repair actions, showing that it can fix different types of bugs. In addition,
\tool fixes bugs that require more repair action than SequenceR, indicating that \tool fixes more complex bugs.} 
}

\subsection{RQ4: Can {\tool} be applied to other programming languages?}

\noindent\textbf{Approach:}
To evaluate the generalizability of \tool to other programming languages, we retrained our models on historical data
for 3 popular programming languages (i.e., Python, C++, and JavaScript). 
For Python, we used Python's version of the QuixBugs benchmark (40 bugs). 
For C++, we evaluate \tool on a popular benchmark
used in previous work~\cite{long2016automatic,long2015staged}. 
Finally, since to the best of our knowledge, there is no benchmark 
for automatic repair in JavaScript, we use the 12 examples associated to common bug patterns in JavaScript described in the Appendix
of previous work~\cite{hanam2016discovering}.
For this RQ, we focus on the output of our neural network, ignoring the validation stage. Since we ignore the validation stage,
we only consider a fix as correct if it's identical to the developer's and consider all generated patches. 

\noindent\textbf{Results:}
\tool performs very well on the Python benchmark, fixing~\pythonfix bugs. 
Figure~\ref{kheapsort} shows a bug that is correctly fixed in the Python dataset. 
\tool fixes \cppfix bugs on the C++ benchmark,. 
We only identified 16 bugs in this benchmark that require modifying a single localization,therefore the majority
of the bugs in the C++ benchmark are out of the scope of \tool.
While \tool performs less well than SPR~\cite{long2015staged}
and Prophet~\cite{long2016automatic} (with 18 and 16 bugs fixed respectively), 
it performs better than Kali~\cite{qi2015analysis} and GenProg~\cite{le2012genprog} (with 2 and 1 bugs fixed respectively). 
In addition, one of the bugs fixed by \tool, \textit{php-309579-309580}, has not been fixed by previous work.
SPR and Prophet performs well on the C++ benchmark but require handcrafted patterns and would need a complete redesign to adapt to another language.
In the JavaScript benchmark, ~\tool fixes \jsfix bugs associated to the ``protect with value check,''
 ``add null place holder,'' and ``propagate to callback'' repair patterns. Figure~\ref{jquery} shows a correct fix by ~\tool for the bug pattern ``protect with value check.''

\begin{figure}
\centering
\begin{lstlisting}[language=Python, basicstyle=\ttfamily\footnotesize, breaklines,escapechar=|,linebackgroundcolor={
	\ifnum \value{lstnumber}=1
	\color{red!10}
	\fi
	\ifnum \value{lstnumber}=2
	\color{green!10}
	\fi
	}]
- for x in arr:	
+ for x in arr[k:]:
\end{lstlisting}
\caption{Correct patch for \textit{KHEAPSORT} Python bug}
\label{kheapsort}
\end{figure}

\begin{figure}
\begin{lstlisting}[language=Java, basicstyle=\ttfamily\footnotesize,  breaklines,linebackgroundcolor={
	\ifnum \value{lstnumber}=1
	\color{red!10}
	\fi
	\ifnum \value{lstnumber}=2
	\color{green!10}
	\fi
	}]
- if (val <= 0) {
+ if (val <= 0 || val == null) {
\end{lstlisting}
\caption{Correct JavaScript patch for a jQuery bug}
\label{jquery}
\end{figure}

\noindent\fbox{
\centering
\parbox{0.95\linewidth}{
\textbf{Summary:} \tool is the first approach that has been successfully applied without major re-implementation to different languages,
fixing a total of \textbf{\alllanguage bugs} in four popular programming languages.
}
}

\subsection{RQ5: Can we explain why {\tool} can (or fail to) generate specific fixes?}

\noindent\textbf{Approach:}
In this RQ, we provide explanations on how \tool generates patches to fix some bugs. 
These explanations are mostly for researchers. We do not provide 
explanations for developers because, unlike classification or defect prediction approaches
that generate a line number which is difficult to analyze, \tool outputs a
complete patch that passes all fault-revealing cases, which the developer can
directly analyze.

\noindent\textbf{The Majority of the Fixes are not Clones:} 
By learning from historical data, the depth of our neural network allows \tool to 
fix complex bugs, including the ones that require generating new variables.
As discussed in Section~\ref{sec:setup}, the evaluation is not valid
if the same bug appears both in the test benchmark and the training or validation sets.
While we removed such cases from our training and validation sets,
the same patch may still be used to fix different bugs introduced at different times in different locations. 
Having such patch clones in 
both training and test sets is valid, as recurring fixes are common.
It is reasonable to and existing static analysis approaches have been proposed to, learn from past fixes to generate fixes for recurring bugs~\cite{meng2011sydit}. 
To understand this effect, we investigate whether the fixes generated by \tool are identical to bug fixes in our training sets.
The majority of the bugs fixed by \tool do not appear in the training sets: only two patches from the C++ benchmark and one from the JavaScript benchmark 
appear in the training or validation sets. This suggests that simple clone-based approaches would fail to generate the fixes, and our NMT-based ensemble learning 
is effective in learning and generating completely different fixes.

\begin{figure}
\begin{lstlisting}[language=Java, basicstyle=\ttfamily\footnotesize,  breaklines,escapechar=|,linebackgroundcolor={
	\ifnum \value{lstnumber}=2
	\color{red!10}
	\fi
	\ifnum \value{lstnumber}=3
	\color{green!10}
	\fi
	}]
// Correct Closure 93 patch generated by ENCORE
- int indexOfDot=namespace.indexOf('.');
+ int indexOfDot=namespace.lastIndexOf('.'); 
\end{lstlisting}
\begin{lstlisting}[language=Java, basicstyle=\ttfamily\footnotesize,  breaklines,escapechar=|,linebackgroundcolor={
	\ifnum \value{lstnumber}=2
	\color{red!10}
	\fi
	\ifnum \value{lstnumber}=3
	\color{green!10}
	\fi
	}]
// Similar change occuring in the training data
- int end = message.indexOf(templateEnd, start );
+ int end = message.lastIndexOf(templateEnd, start ); 
\end{lstlisting}
\caption{Closure 93 patch and similar training data change}
\label{closure93}
\end{figure}

\noindent\textbf{\textit{Closure 93} from Defects4J:}
\textit{Closure 93} in the Defects4J benchmark is one of the bugs only \tool can fix and is displayed in Figure~\ref{closure93}.
The bug is fixed by replacing the method call \texttt{indexOf} with the method \texttt{lastIndexOf}. \tool is able to make this correct change 
because of a similar change in our training set (in the hibernate-orm project,
also shown in Figure~\ref{closure93}).
However, the change is not a simple clone since all the variable names are completely different, which indicates that \tool can learn some abstraction of the variable names
in the original statement.

\noindent\textbf{\textit{Lang 26} from Defects4J:}
\tool can also fix bugs that require more complex changes.
For example, the fix for \textit{Lang 26} from the Defects4J dataset shown in Figure~\ref{lang26} requires injecting a new variable \texttt{mLocale}.
This new variable only appears four times in our training set, and never in a similar context. \tool is still able to generate the correct fix
because, thanks to our tokenization approach, \texttt{mLocale} is divided into the tokens \texttt{m} and \texttt{Locale}.
The token \texttt{Locale} occurs in our training set in similar context to the tokens \texttt{Gregorian}, \texttt{Time}, and \texttt{Zone} which are 
all in the buggy statement. 

\begin{figure}[tb]
	\centering
	\includegraphics[width=0.49\textwidth]{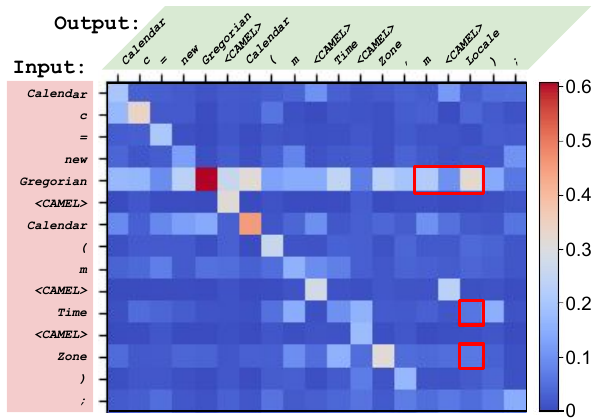}
	\caption{Attention map for the correct patch of \textit{Lang 26} from the Defects4J benchmark generated by~\tool}
	\label{fig:attlang26}
\end{figure}

The attention map in Figure~\ref{fig:attlang26} confirms that these three tokens are important for generating the \texttt{Locale} variable.
Specifically, the tokenized input is shown on the y-axis while the tokenized generated output is displayed on the x-axis. 
The token \texttt{<CAMEL>} between \texttt{m} and \texttt{Locale} indicates that these two tokens form one unique variable.
The attention map shows the relationship between the input token (vertical axis) and the generated tokens (horizontal axis).
The color in a cell represents the relationship of corresponding input and output tokens with red colors indicating that an input token
is influential in generating the output token, while darker blue colors indicate that the input token has little to no influence on the output token.
For example, to generate the token \texttt{Locale}, the most influential input token is \texttt{Gregorian} (in light red), followed by \texttt{Time} and \texttt{Zone} (in light blue). On
the other hand, the \texttt{c} token (in dark blue) does not influence the generation of \texttt{Locale}.

This example shows that \tool can generate patches, even if the exact same pattern has not occurred in the training set. For \textit{Lang 26}, \tool generates a 
correct variable name, that never appears in the same context during training.

\begin{figure}
\begin{lstlisting}[language=Java, basicstyle=\ttfamily\footnotesize, breaklines,escapechar=|,linebackgroundcolor={
	\ifnum \value{lstnumber}=2
	\color{red!10}
	\fi
	\ifnum \value{lstnumber}=3
	\color{green!10}
	\fi
	}]
// Correct Lang 26 patch generated by ENCORE
- Calendar c = new GregorianCalendar(mTimeZone);
+ Calendar c = new GregorianCalendar(mTimeZone, mLocale);
\end{lstlisting}
\begin{lstlisting}[language=Java, basicstyle=\ttfamily\footnotesize, breaklines,escapechar=|,linebackgroundcolor={
	\ifnum \value{lstnumber}=2
	\color{red!10}
	\fi
	\ifnum \value{lstnumber}=3
	\color{green!10}
	\fi
	\ifnum \value{lstnumber}=4
	\color{green!10}
	\fi
	}]
// Similar change occuring in the trainning data
- Calendar calendar = new GregorianCalendar();      
+ Calendar calendar = new GregorianCalendar(
           TimeZone.getDefault(), Locale.getDefault());
\end{lstlisting}
\caption{Lang 26 patch and similar training data change}
\label{lang26}
\end{figure}

\begin{figure}
\begin{lstlisting}[language=Java, basicstyle=\ttfamily\footnotesize, breaklines,escapechar=|,linebackgroundcolor={
	\ifnum \value{lstnumber}=2
	\color{red!10}
	\fi
	\ifnum \value{lstnumber}=3
	\color{green!10}
	\fi
	}]
// Correct BUCKETSORT patch generated by ENCORE
- for ( Integer count : arr ) {   
+ for ( Integer count : counts ) { 
\end{lstlisting}
\begin{lstlisting}[language=Java, basicstyle=\ttfamily\footnotesize, breaklines,escapechar=|,linebackgroundcolor={
	\ifnum \value{lstnumber}=2
	\color{red!10}
	\fi
	\ifnum \value{lstnumber}=3
	\color{green!10}
	\fi
	}]
// Correct Lang 59 patch generated by ENCORE
- str.getChars(0, strLen, buffer, size);
+ str.getChars(0, width, buffer, size);
\end{lstlisting}
\caption{Patches that requires replacing variable}
\label{bucket}
\end{figure}

\noindent\textbf{\textit{BUCKETSORT} from QuixBugs and \textit{Lang 59} from Defects4J:}
Figure~\ref{bucket} shows two examples (\textit{BUCKETSORT} from QuixBugs and \textit{Lang 59} from Defects4J) of correct patches
generated by {\tool} that require replacing a variable name by another one.
These two bugs can only be fixed by the fourth model of our ensemble learning approach, which indicates that this specific model might be more successful
than the others in learning this specific pattern.

\noindent\textbf{\textit{SUBSEQUENCE} from Quixbug:}
Figure~\ref{subseq} shows an overfitted patch---an incorrect patch that makes the tests pass---generated by \tool.
Our overfitted patch returns \texttt{Arrays.asList}, which is an \texttt{Array.ArrayList} (which has fixed length according to the JavaDoc)
while the correct patch returns a normal \texttt{ArrayList}.
At first glance, the patch that returns a fixed length \texttt{Array.Array\-List} should not pass
the test cases because the algorithm needs to add elements to this \texttt{Array.ArrayList} later.
After further investigation, we found that when an \texttt{ArrayList}
is used as input to \texttt{asList}, the returned \texttt{Array.ArrayList} is backed by that \texttt{Array\-List} and does not have fixed length,
which allows our patch to pass all the test cases.
This overfitted patch highlights a potential bug in the \texttt{asList}
method of the \texttt{java.util} library since the return value of \texttt{asList}
does not behave as expected. This shows \tool also has the potential  
of generating mutants that can find bugs in software.

\begin{figure}
\begin{lstlisting}[language=Java, basicstyle=\ttfamily\footnotesize, breaklines,escapechar=|,linebackgroundcolor={
	\ifnum \value{lstnumber}=2
	\color{red!10}
	\fi
	\ifnum \value{lstnumber}=3
	\color{green!10}
	\fi
	}]
// Overfitted patch generated by ENCORE
- return new ArrayList();
+ return new ArrayList<>(Arrays.asList(new ArrayList()));
\end{lstlisting}
\begin{lstlisting}[language=Java, basicstyle=\ttfamily\footnotesize, breaklines,escapechar=|,linebackgroundcolor={
	\ifnum \value{lstnumber}=2
	\color{red!10}
	\fi
	\ifnum \value{lstnumber}=3
	\color{green!10}
	\fi
	}]
// Correct Patch
- return new ArrayList();
+ return new ArrayList<>(new ArrayList()));
\end{lstlisting}
\caption{Overfitted patch for SUBSEQUENCE}
\label{subseq}
\end{figure}
\section{Limitations} 
There is randomness in the training process of deep learning model.
We perform multiple runs and find that the randomness in training has little impact on the performances of the final trained models.
The hyper-parameters tuning process also contains randomness. We tuned our model 
for five days, investigating almost 400 different sets of hyper-parameters 
using random search (Section~\ref{sec:ensemble}).
A main challenge of deep learning is to explain the output of a neural network.
Fortunately, for developers, the repaired program that compiles and passes test cases should be self-explanatory.
For users who build and improve \tool models, we leverage the recent multi-step attention mechanism~\cite{gehring2017convolutional} to explain why a fix was generated or not. 
For a complete end-to-end automatic program repair solution, the buggy line localization is needed.
While we assume perfect localization, 
we also show that for all except two of the bugs that \tool fixes, their localized lines can be identified.

\section{Related Work}

\noindent\textbf{Deep Learning for APR:}
Deep learning has been used to fix compilation errors~\cite{gupta2017deepfix}. More recently, NMT has been used for 
automatic program repair~\cite{tufano2018empirical,devlin2017semantic,chen2018sequencer}.
Devlin et al. work~\cite{devlin2017semantic} used a neural network architecture to repair bugs in Python,
focusing on four common transformations. They evaluated their model by injecting bugs generated from the 
same four transformations. While their model performed relatively well on injected bugs, it is limited as it can
only fix bugs that follow the specific types of transformations and it is unclear whether it would work for fixing real-world bugs.
A similar approach using an LSTM-based NMT was investigated by Tufano et al.~\cite{tufano2018empirical}; however,
this approach is limited to fixing bugs inside small methods and only generates a template of the fix,
without generating the full correct statements. SequenceR\cite{chen2018sequencer} is a concurrent work
that includes patch validation.
Compared to SequenceR, \tool uses a different deep learning model combined with ensemble learning that fixes more bugs.
In addition, \tool is generalizable to different programming languages while SequenceR focused on fixing
Java bugs.

\noindent\textbf{\gv Program Repair:}
Many APR techniques have been proposed~\cite{le2012genprog,long2015staged,saha2017elixir,xuan2017nopol,martinez2016astor,long2017automatic,xin2017leveraging,xiong2017precise,jiang2018shaping,hua2018sketchfix,wen2018context,liu2018mining,chen2017contract,le2016history}. 
We use a completely different approach compared to these techniques, and as shown in Section~\ref{eval}, our approach fixes  bugs that 
existing techniques have not fixed. 
In addition, these techniques require significant domain knowledge and manually crafted rules that are language dependent, while 
thanks to our ensemble NMT approach, \tool automatically learns such patterns and is generalizable to several programming languages with minimal effort.

\noindent\textbf{Grammatical Error Correction:}
The counterpart of automatic program repair in NLP is grammatical error correction (GEC).
Recently, work in the field of GEC mostly focuses on using machine translation in fixing grammatical 
errors~\cite{chollampatt2018mlconv,Liu2017,Schmaltz2017,Napoles2017,Sakaguchi2017,Chollampatt2018,Junczys-Dowmunt2018,Yannakoudakis2017,Ge2018,Ge2018a,Kaneko2017}.
Recent work~\cite{chollampatt2018mlconv} applied an attention-based convolutional encoder-decoder model to correct sentence-level grammatical errors.
\tool{} is a new application of NMT models on source code and programming languages, which addresses unique challenges.

\noindent\textbf{Deep learning in software engineering:}
The software engineering community had applied deep learning to performing 
various tasks such as defects prediction~\cite{wang2016automatically,li2017software,wang2017software},
source code representation~\cite{alon2019code2vec,wang2018dynamic,allamanis2018learning,peng2015building},
source code summarization~\cite{gu2018deep,allamanis2016convolutional}
source code modeling~\cite{white2015toward,hellendoorn2017deep,allamanis2017survey,chakraborty2018tree2tree},
code clone detection~\cite{white2016deep,li2017cclearner}, and
program synthesis~\cite{murali2018neural,ling2016latent,mou2015end,alexandru2016guided}.
Our work uses a new deep learning approach for automated program repair. 
\section{Conclusion}

We propose \tool, a new end-to-end approach using NMT and ensemble learning to automatically repair 
bugs in multiple languages.
We evaluate \tool on five benchmarks in four different programming languages and
found that \tool can repair \alllanguage{} bugs 
including \uniquejava
that have not been fixed before by existing techniques.
In the future, we plan to improve our approach to
work on multi-localization bugs and find an effective way to represent the context
of a bug.

\bibliographystyle{unsrt}  
\bibliography{paper} 

\end{document}